\documentclass[12pt]{article}

\usepackage{epsfig}
\usepackage{amsmath}
\usepackage{amssymb}
\begin{document}

\title{Mean first passage time for fission potentials having structure}
\author{H. Hofmann $^{1}$ and A.G. Magner$^{1,2}$\\ \small\it{1)
Physik-Department, TU M\"unchen, 85747 Garching, Germany } \\
\small\it{2) Institute for Nuclear Research, 03028 Kiev-28,
Ukraine}}
 \maketitle

\begin{abstract}
A schematic model of over-damped motion is presented which permits
one to calculate the mean first passage time for nuclear fission.
Its asymptotic value may exceed considerably the lifetime
suggested by Kramers rate formula, which applies only to very
special, favorable potentials and temperatures. The additional
time obtained in the more general case is seen to allow for a
considerable increment in the emission of light particles.
\end{abstract}

\begin{center}PACS numbers: 24.75.+i, 24.60.Dr, 24.10.Pa, 24.60.-k
\end{center}

Fission experiments are commonly analyzed with the help of
statistical codes which are based on simple rate formulas for the
processes of fission and emission of light particles. In the early
80'ies an excess of neutrons was observed over that for which the
fission rate is simply estimated by the Bohr Wheeler formula
$\Gamma_{\text{f}}\equiv \Gamma_{\text{BW}}$ (see e.g. the review
articles \cite{HILSCHER, pauthoe} with references to original
work). An improvement was seen in replacing $\Gamma_{\text{BW}}$
by the $\Gamma_{\text{K}}$ of Kramers \cite{kram} in which the
fission rate formula gets reduced by friction, the more the larger
the dissipation strength. Additional possibilities for enhancing
the relative emission probability
$\Gamma_{\text{n}}/\Gamma_{\text{f}}$ of light particles like
neutrons were attributed to two effects which seem to arise in a
time dependent description: (i) Starting the dynamics of fission
at some time zero, it takes a finite time for the current across
the barrier to reach thestationary value from which Kramers
derived his formula. (ii) To this stationary current a finite time
lapse $\tau_{\text{ssc}}$ may be associated for the motion from
the saddle point down to scission. Often feature (i) is
interpreted as a delay of fission during which particles may be
emitted on top of the number given by
$\Gamma_{\text{n}}/\Gamma_{\text{K}}$. Likewise, it is believed
that also the neutrons emitted during $\tau_{\text{ssc}}$ are not
accounted for by this $\Gamma_{\text{n}}/\Gamma_{\text{K}}$.

A review of these features and of their practical applications can
be found in \cite{pauthoe}. It can be said that interesting
consequences have been deduced in this way, both for the value of
the dissipation strength as well as for its variation with
temperature, see e.g. \cite{HBDMSVP, dioszegi}. More recently,
however, the question has been raised as to whether Kramers'
original rate formula itself accounts for realistic situations in
fission \cite{hiry}. For under-damped motion, modification becomes
necessary whenever the inertia changes from the minimum to the
barrier. Moreover, it has been argued that any temperature
dependence of the pre-factor must not only be attributed to
friction, but also to the inertia and in particular to the
stiffnesses of the potential at its extrema. In
\cite{HoIv-MFPT-PRL} the interpretation of fission decay as a
sequence of three subsequent steps (minimum-saddle, motion across
saddle, saddle-scission) has been re-examined with the help of the
concept of the "mean first passage time" (MFPT). Restricting to
over-damped motion such an analysis can be performed in analytic
fashion, simply because for this case an analytic formula for the
time $\tau_{\text{mfpt}}$ exists. It reads
\begin{equation}\label{MFPT-smo} \tau_{\text{mfpt}}(Q_{\text{a}} \to Q_{\text{ex}})=
\frac{\gamma}{T}\int_{Q_{\text{a}}}^{Q_{\text{ex}}} du
\;\exp\left[\frac{V(u)}{T}\right] \int_{-\infty}^u dv
\;\exp\left[-\frac{V(v)}{T}\right]\,,\end{equation}and is valid if
any coordinate dependence of friction $\gamma$  and temperature
$T$ are discarded, details may be found in \cite{gardiner-STM}.
Here, $Q_{\text{a}}$ is meant to represent that minimum of the
potential $V(Q)$ which is associated to the "ground state
deformation" and $Q_{\text{ex}}$ stands for the "exit point". In
this sense the $\tau_{\text{mfpt}}(Q_{\text{a}} \to
Q_{\text{ex}})$ determines the average time the system spends in
the interval from $Q_{\text{a}}$ to $Q_{\text{ex}}$. It is
calculated for a situation where the system, after starting at
$Q_{\text{a}}$ sharp, does not return to this interval once it has
crossed the point $Q_{\text{ex}}$, which is then referred to as an
"absorbing barrier". Typical for fission, for $Q\to -\infty$ the
$V(Q)$ is assumed to rise to plus infinity, and hence acts as a
"reflecting barrier".
\begin{figure}
\begin{center}
\epsfig{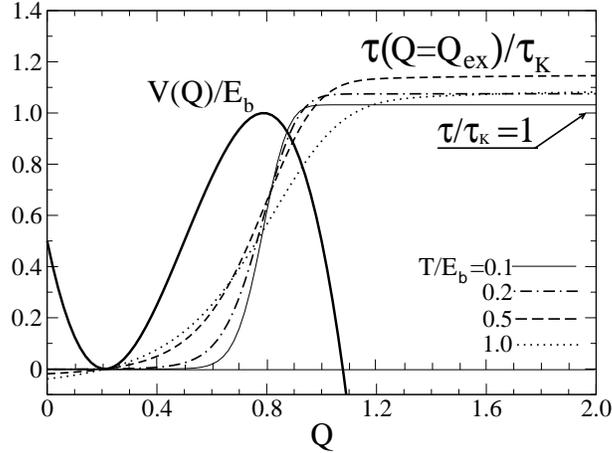}
\caption{The mean first passage time $\tau(Q=Q_{\text{ex}})\equiv
\tau_{\text{mfpt}}(Q_{\text{a}} \to Q_{\text{ex}})$ of
(\ref{MFPT-smo}) (normalized to the $\tau_{\text{K}}$ of
(\ref{tau-smol})) for a cubic potential and different
 temperatures, defined in units of $E_{\text{b}}$.}
  \label{cubic}
\end{center}
\end{figure}

As will be demonstrated again below, the
$\tau_{\text{mfpt}}(Q_{\text{a}} \to Q_{\text{ex}})$ tends to a
constant as soon as the exit point is sufficiently far to the
right of the potential barrier. This constant, which henceforth
shall simply be called $\tau_{\text{mfpt}}$, becomes identical to
the inverse of  Kramers' rate $\tau_{\text{mfpt}} =
\tau_{\text{K}} \equiv \hbar/\Gamma_{\text{K}}$, whenever the
usual conditions are fulfilled which underly Kramers' derivation.
Recalling that we are dealing with over-damped motion, this
$\tau_{\text{K}}$ is given by
\begin{equation}\label{tau-smol}\tau_{\text{K}} = \frac{2\pi \gamma}
{\sqrt{C_{\text{a}}|C_{\text{b}}|}}\exp(E_{\text{b}}/T)\,,
\end{equation} where $C_{\text{a}}$ and $C_{\text{b}}$ are the stiffnesses
at the potential minimum and barrier, respectively. In
\cite{gardiner-STM} this fact is proven by applying the saddle
point approximation to formula (\ref{MFPT-smo}). For this it is
important to have exactly {\em two} saddle points, those
corresponding to {\em one} minimum and {\em one} barrier. Notice
that the saddle point approximation requires one to replace the
barrier by an inverted oscillator, which indeed was also assumed
to hold true by Kramers in his famous work.

As a typical case we show in Fig.\ref{cubic} the results of
calculations of $\tau_{\text{mfpt}}(Q_{\text{a}} \to
Q_{\text{ex}})/\tau_{\text{K}}$ for the same cubic potential as
used in \cite{HoIv-MFPT-PRL}. It may be specified by its first
derivative to be given by the form
\begin{equation}\label{pot-cubic} V^{\,\prime}(Q) \,\propto\,
\left(Q-Q_{\text{a}}\right)\left(Q-Q_{\text{b}}\right)\,,
\end{equation} with the barrier height $E_{\text{b}} =
V(Q_{\text{b}})-V(Q_{\text{a}})$ to be $8$ MeV with the extrema to
lie at $Q_{\text{a}}\simeq 0.2$ and $Q_{\text{b}}\simeq  0.8$. It
is observed that the asymptotic ratio
$\tau_{\text{mfpt}}/\tau_{\text{K}}$ becomes close to unity,
indeed, if only the parameter temperature over barrier height
becomes small enough. However, even for this case of exactly two
well pronounced extrema, deviations from unity are clearly visible
at larger temperatures.

\begin{figure}[ht]
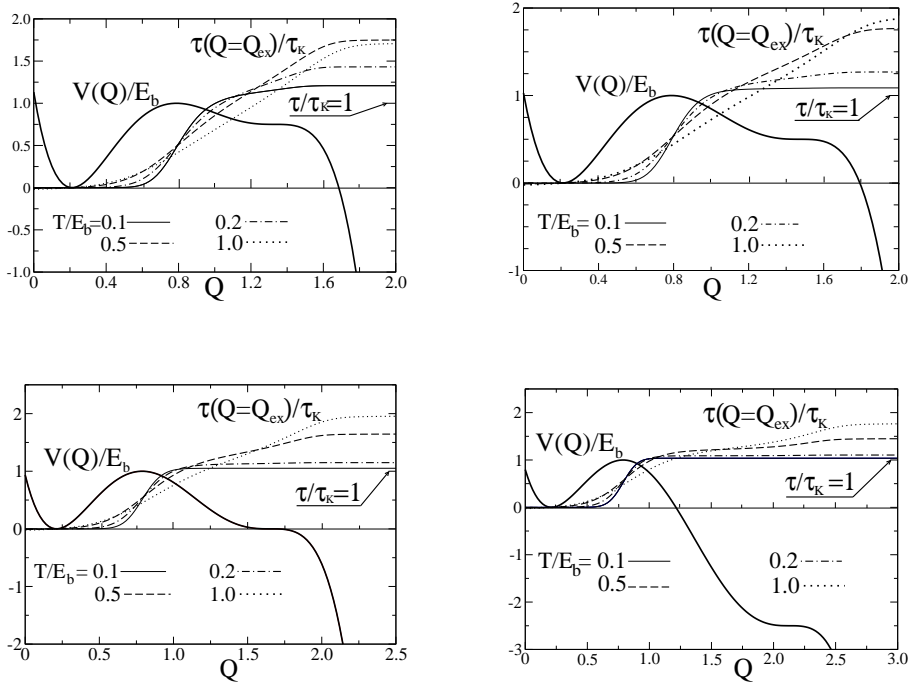

\vspace{10pt}
\begin{center}
 \epsfig{figure=t5_6.eps,width=0.35\linewidth}
 \hspace{30 pt}
 \epsfig{figure=t5_4.eps,  width=0.35\linewidth}
\\[30pt]
 \epsfig{figure=t5_0.eps,   width=0.35\linewidth}
 \hspace{30 pt}
 \epsfig{figure=t5_m20.eps, width=0.35\linewidth}
\caption{\label{poly5-shoul} Same as in Fig.\ref{cubic} but for
potentials having shoulders at some
 $Q_{\text{s}}$ of heights $V_{\text{s}}/E_{\text{b}}$
  relative to the barrier: $V_{\text{s}}/E_{\text{b}}=0.75$
  top left, $0.5$ top right, $0$ bottom left, $-2.5$ bottom right.}
\end{center} \end{figure}
This situation becomes more dramatic as soon as the potential
shows additional structure. This will now be demonstrated using a
schematic potential of fifth order in $Q$. Again, $V(Q)$ will be
fixed by its first derivative,
\begin{equation}\label{pot-poly5} V^{\,\prime}(Q) \,\propto
\left(Q-Q_{\text{a}}\right)\left(Q-Q_{\text{b}}\right)
\left(Q-Q_{\text{c}}\right)\left(Q-Q_{\text{d}}\right)\,,
\end{equation} with $E_{\text{b}}$, $Q_{\text{a}}$  and
$Q_{\text{b}}$ unchanged. The remaining two parameters
$Q_{\text{c}}$ and $Q_{\text{d}}$ may be used to specify structure
of the potential beyond the barrier. In Figs.\ref{poly5-shoul}
they have been chosen to be identical to one another,
$Q_{\text{c}}=Q_{\text{d}}= Q_{\text{s}}$, with their values fixed
such that the height $V_{\text{s}}$ of the then existing shoulder
takes on the values specified in the figure captions.

In all cases the calculation of $\tau_{\text{mfpt}}(Q_{\text{a}}
\to Q_{\text{ex}})$ was performed up to regions of the exit point
$Q_{\text{ex}}$ where the stationary value is reached. It is seen
that this asymptotic regime is not very far away from the one
where the potential is assumed to have additional structure. This
is so even for the example shown in the lower right corner of
Fig.\ref{poly5-shoul}. There a potential is taken with a shoulder
in a region which lies $20$ MeV below the first minimum, or $-2.5
E_{\text{b}}$ in terms of the barrier height. For heavy nuclei
this may thus be said to correspond to the scission region after
which the fragments separate. The ratio
$\tau_{\text{mfpt}}(Q_{\text{a}} \to
Q_{\text{ex}})/\tau_{\text{K}}$ shown in the figures is calculated
for the $\tau_{\text{K}}$ of (\ref{tau-smol}). Evidently friction
drops out but the stiffnesses $C_{\text{a}}$ and $C_{\text{b}}$
from (\ref{tau-smol}) remain. They are taken to be those of the
individual potentials for which the $\tau_{\text{mfpt}}$ is
computed. These results exhibit clearly the mistake one makes if
only Kramers' rate formula is used to estimate the time the system
stays together. Rather, the considerable overshoot of
$\tau_{\text{mfpt}}$ over $\tau_{\text{K}}$ indicates that much
more time is available for light particles to be emitted before
scission. Suppose we look at neutrons. Whenever, their average
width $\Gamma_{\text{n}}$ may be used to calculate their
multiplicity per fission event from $\Gamma_{\text{n}}/
\Gamma_{\text{f}}$ the enhancement of this number over that given
by $\Gamma_{\text{n}}/\Gamma_{\text{K}}$ is determined by the
ratio $\tau_{\text{mfpt}}/\tau_{\text{K}}$, viz
\begin{equation}\label{part-widt-nf}\frac{\Gamma_{\text{n}}}
{\Gamma_{\text{f}}} = \frac{\Gamma_{\text{n}}} {\Gamma_{\text{K}}}
\frac{\tau_{\text{mfpt}}}{\tau_{\text{K}}}\,.\end{equation} For
the potentials chosen here this enhancement may become quite
large.

\begin{figure}
\begin{center}
\epsfig{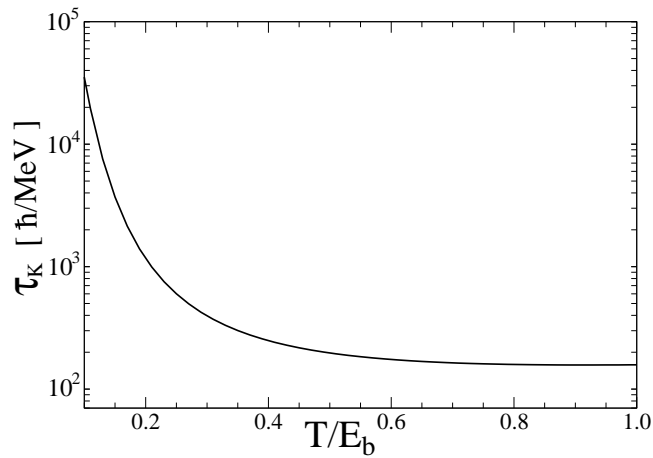}
 \caption{The MFPT corresponding to Kramers' estimate as function of
 temperature in units of the barrier height, see text.}
  \label{tktcol}
\end{center}
\end{figure}
To get some feeling for absolute values of this extra available
time we estimated the pre-factor
$\gamma/\sqrt{C_{\text{a}}|C_{\text{b}}|}$ of (\ref{tau-smol})
following suggestions given in \cite{hiry}. We simply replaced the
geometric mean $\sqrt{C_{\text{a}}|C_{\text{b}}|}$ by the $C$
which there in sect.III.B.5 appears in the relaxation time
$\tau_{\text{coll}}$ for over-damped collective motion. Rather
than use the formulas given in that section we simply take the
results shown in Fig.4 (by the dashed line). Between temperatures
of one and four MeV this $\tau_{\text{coll}}$ shows an almost
linear dependence in $T$ such that one may write
\begin{equation}\label{avtaucoll}\frac{\gamma}
{\sqrt{C_{\text{a}}|C_{\text{b}}|}} =
\overline{\tau_{\text{coll}}} \;\simeq \;- \frac{3}{4} +
\frac{5}{4}\, T
\;\left(\frac{\hbar}{\text{MeV}}\right)\,.\end{equation} Putting
this estimate into the formula given in (\ref{tau-smol}) for
$\tau_{\text{K}}(T)$ one obtains the curve shown in
Fig.\ref{tktcol}. The strong temperature dependence reflects the
exponential function $\exp(E_{\text{b}}/T)$. As the estimate
(\ref{avtaucoll}) seizes to be valid above $T\simeq 4$ MeV the
curve should not be taken too seriously above
$T/E_{\text{b}}\simeq 0.5$. For such a $T$ the $\tau_{\text{K}}$
is about $200\,\hbar /\text{MeV}$ large. As the
$\tau_{\text{mfpt}}/\tau_{\text{K}}$  typically is about $1.5$ the
{\em additional time increment}
$\Delta\tau_{\text{f}}=\tau_{\text{mfpt}} - \tau_{\text{K}}$ takes
on  the sizable value of roughly $100\,\hbar /\text{MeV}$, and,
hence, is at least as large as a typical transient time.

\begin{figure}
\begin{center}
\epsfig{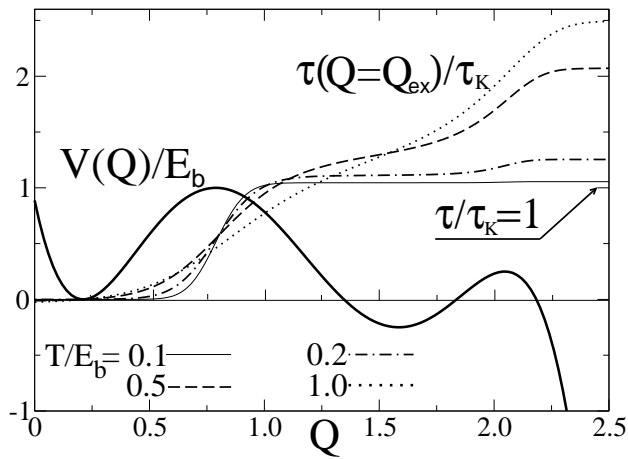}
 \caption{Like in Fig.\ref{poly5-shoul} but for a potential having an
 additional minimum of depth $V_{\text{c}} = - 0.25\,E_{\text{b}}$ and
 a second maximum of height $V_{\text{d}} = 0.25\,E_{\text{b}}$.}
  \label{t5_m2_2}
\end{center}
\end{figure}

Finally, in Fig.\ref{t5_m2_2} we look at the case of the potential
having a second minimum and maximum. As was to be expected the
effect is even larger than before. We would like to remark,
however, that this example should be taken with some caution.
Commonly, such a double humped barrier comes about because of
shell effects. In the range of temperatures considered here the
latter may be considered to be quite weak if not already washed
out completely. After all our study is concerned with over-damped
motion. According to \cite{hiry} (see also \cite{hofrep}) nuclear
collective motion may be expected to become over-damped only above
temperatures of about $T=2$ MeV.

Our results may be summarized as follows. One of the main issues
has been to corroborate features suggested before in
\cite{HoIv-MFPT-PRL}, and to some extent already in \cite{hiry}.
In essence they imply the following two issues: (i) For situations
for which transport equations like those of Kramers or
Smoluchowski (or the corresponding Langevin equations) may be
applied to analyze fission experiments, it may be inadequate to
simply use Kramers' famous rate formula. Deviations from that may
originate in various reasons, for instance in transport
coefficients varying with shape, see \cite{hiry}. Here, we
concentrated on properties of the potential restricting ourselves
to over-damped motion and the model case of constant friction.
(ii) For this model we have been able to demonstrate that there is
considerable room for increasing the time the fissioning system
stays together without having to rely on concepts which are
meaningful only within a time dependent picture. Whereas results
obtained within the latter may depend crucially on initial
conditions this is not the case for the MFPT \cite{HoIv-MFPT-PRL}.
This $\tau_{\text{mfpt}}$ represents the average time it takes for
the system to start at the potential minimum and to make its
motion all the way out to scission. It includes relaxation
processes around the first minimum as well as the sliding down
from saddle to scission. The way it is derived \cite{gardiner-STM}
implies a proper incorporation of averages over the statistics
which are to be associated with a process underlying fluctuating
forces. Whereas for Kramers' model case the $\tau_{\text{mfpt}}$
is nothing else but his inverse rate, this is no longer true for
larger temperatures and, in particular, not for potentials of more
complicated structure. We have been able to demonstrate that the
latter may lead to a considerable prolongation of the time the
systems spends before it scissions, allowing in this way for
emission of light particles on top of those given by
$\Gamma_{\text{n}}/\Gamma_{\text{K}}$. Of course, further work
will be necessary to clarify the relevance of this feature with
respect to real situations. For such studies not only more
realistic potentials have to be used, one should also try to
generalize formula (\ref{MFPT-smo}) to include a coordinate
dependent friction coefficient. The ultimate goal should be to be
able to use such a formula in a statistical code where one may
account for temperatures changing during the process.

{\em Acknowledgements:} The authors wish to thank R. Hilton, F.A.
Ivanyuk and C. Rummel for useful help. They also wish to
acknowledge financial support from the "Deutsche
Forschungsgemeinschaft". One of us (A.G. Magner) would like to
thank the Physik Department of the TUM for the kind hospitality
extended to him.

\end{document}